# Progress in the short-term earthquake forecasting by integrating KaY wave analysis and satellite observations

D.Ouzounov,[1, 3, *] A.Yagodin,[2, 3]

*[1]Center of Excellence in Earth Systems Modeling
& Observations (CEESMO), Chapman University
Orange, CA, USA*
*[2]Laboratory for earthquake prediction, Haifa, Israel*
*[3]Ertha Technologies, Herndon, VA, USA*

Following N.Kozyrev's idea about the influence of the gravitational fields of the Sun and the Moon on the Earth's crust, we consider a low-frequency resonance of the Earth's crust blocks is happening before the earthquakes occurrence. The resonance affects several geophysical parameters and is associated with many short-term precursory phenomena; the most notable are the release of underground gases (radon), the air temperature rises, change of thermal radiation in the troposphere, and an increase in the electron density in the ionosphere. During the final stage of the earthquake genesis, long-period gravity-seismic waves called the Kozyrev-Yagodin ( KaY-) wave are formed and moves from the periphery to the epicenter of the future earthquake. We discus that by integrating information from different stages of earthquake genesis: the network observations of KaY waves and satellite thermal radiation, we significantly could advance the accuracy and reliability of short-term earthquake forecasting, previously unachievable.

PACS numbers: 07.57.-c;42.68Wt;42.68Ay; 91.30Px

## I. INTRODUCTION

The influence of short-term lithospheric processes on geophysical parameters associated with the occurrence of earthquakes has been studied over the last 30 years and led to the creation of lithosphere-atmosphere-ionosphere coupling (LAIC) understanding about pre-earthquake processes [1,2,3,4,5,6].In - mid - 90s it was established that all anomalous variations of different parameters presented are observed within the earthquake preparation / area of activation zone, which is correlated with the magnitude of an impending earthquake. Despite the progress in understanding the underlying processes, the initiating of short-term pre-earthquake phase is still unknown. Currently many scientists are studying the influence of solar activities, Moon phases and geomagnetic storms on seismic activity. In this paper we are exploring further the initial Kozyrev's idea about the Earth's synchronicity and the Moon's tectonics [7] . We consider that under the influence of the gravitational fields of the Sun and the Moon a low-frequency resonance was initiated inside the Earth's crust blocks [8,9,10]. The resonance of the Earth's crust is associated with the release of underground gases radon, electromagnetic (EM) emissions, a rise in temperature in the atmosphere, and an increase/decrease in electron density in the ionosphere. Thus, the coupling between the lithosphere, the atmosphere and the ionosphere, is initiated [5].

We are presenting a joint study, of KaY waves (a long-period gravity-seismic waves) and satellite thermal radiation capable to trace the finale genesis of the earthquake. The main concept behind the joint use was to follow the earthquake development during the final two weeks with different methods of spatial and temporal monitoring that can provide forecasting about the final seismic rupture initiation hours/ten of hours in advance.

## II. SEISMO GRAVITY OBSERVATIONS. KaY WAVES

In 1972 N.Kozyrev presented a hypothesis about the relationship between the tectonic processes of the Earth and the Moon [7]. He has shown the existence of a trigger of tidal influences through the gravitational effects of the Earth and the Moon. The existence of a direct causal connection between the tectonic processes of the Earth and the Moon is still not fully understood. Perhaps this is due to the synchronism of the mountain-building cycles of the Earth and the Moon and to the gravitational interaction of volumetric bodies.

Gravitational seismic Resonance (GSR) means an increase in amplitude oscillations of the resonator under external influence (Solar and Moon and other planetary forces) at the coincidence of the natural frequency of the resonator and the resonant external force frequency [11]. For the earthquake preparation area, under the resonator area, we can accept the area in the lithosphere where at least one earthquake happened. Potentially earthquakes resonators may be still undetected that have not been included in statistical bases of earthquakes . In the region of the earthquake preparations, appear multiple earthquake indicators detected by instrumental methods during the time of the GSR .In principle the pre-earthquake signals such as an increase in the flow of radon and carbon dioxide, EM signals, atmospheric temperature anomalies, ionospheric anomalies, are indicators that a cluster of earthquake type of resonators begins to show emergent properties due to continued external influence on the system that has reached the critical level [11]. We have found that the resonance occurs at the site of the future epicenter (hypocenter) of the earthquake, then it develops in depth until the end of the resonance or until it reaches the surface (at its maximum) on the opposite side of the planet , see Fig.1 [12]. Then a reverse gravitational-seismic body wave appears, the projection of which on the surface moves at a speed of about 100 km / h. It returns to the place where the resonance began and causes explosive destruction in it - an earthquake that lasts about 20 seconds. The return long-period gravity-seismic waves with a speed about 100 km / h is called KaY-wave ( Kozyrev-Yagodin wave) are formed in the final stages of the genesis of an earthquake cycle. During the experimental observations of gravity-seismic waves, the analysis of the connection between the arising pre-earthquake signals and the physical processes occurring in the Earth's crust, revealed that the KaY-wave arises at the last stage and moves from the periphery to place location of epicenter of the future earthquake. During the field observation of KaY waves, a cloud pattern was discovered and described with a special shape (Heralds), indicating for the presence of low-frequency oscillations with the formation of a standing wave (Fig.2). Such waves are emitted by the earth's crust and are projected upward, as if on a screen by condensation of the water vapor. These waves create crests of harmonic oscillations elongated along the wave front with a perpendicular to the side of the epicenter of the future earthquake.[8,10]. The presence of a standing wave projection indicated that these oscillations have the character of a wave moving towards the epicenter of a future earthquake. Wave speed of KaY is visible from the correlation with the earthquake events.

*Electronic address: dim.ouzounov@gmail.com

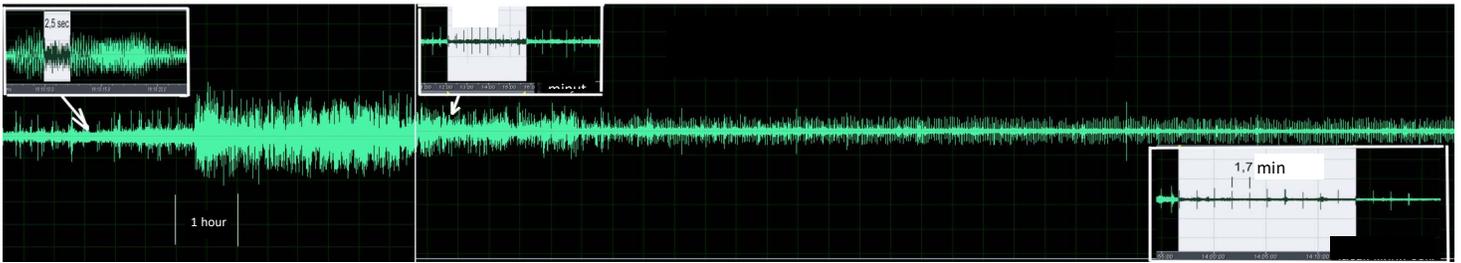

Fig.1 Example of recording the Seismo gravity (SG) resonance on a sensor in Haifa (Israel) of resonant oscillations in the Dodecanese islands (Greece).

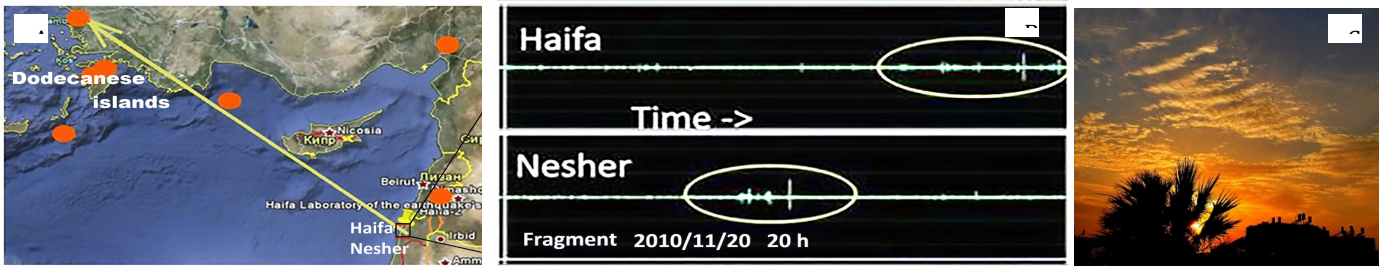

Fig.2 Example of recording the SG resonance A/ and B/ Comparison of the sequence of registration of anomalies on station sensors when the wave moves towards Dodecanese islands. Wave peaks are first recorded at the station in Nesher and then in Haifa.; C/ Photo of precursor clouds with a perpendicular line pointing towards Greece. (Reminder: these are not earthquake peaks, but peaks of KaY wave recorded by the sensors in Haifa and Nesher , 7 hours before the earthquake that occurred at the in the Dodecanese islands )

The first operational testing for using KaY observations for short-term forecasting was performed in 2012 [9]. To the Earthquake Expert Council of Russia short-term forecasting of earthquakes was carried out generating hours/tens of hours of warning before the start of the earthquake, based on monitoring the KaY-wave.

The Russian Emergencies Ministry coordinated the tests. During the tests, about a dozen warnings were issued for earthquakes in the Mediterranean Sea, the Middle East, Russia, Chile, without a single false alarm. (See the supplementary materials for additional information).

## III. METHOD OF SATELLITE RADIATION ANOMALY

The search for recent data analysis from several space-based instruments and ground-based observations has provided evidence of the existence of atmospheric signals before the earthquake [3,14]. These studies have expanded our understanding of earthquakes' physics and the phenomena that precede their energy release. Studies of the relationship between satellite thermal infrared (TIR) data and earthquakes have been based on single and multi-instrument recordings. Advanced Very High Resolution Radiometer (AVHRR) images have been used to develop simple analysis methods based on comparing images before and after at the epicenter of an earthquake [15, 16]. The latest satellite observations make it possible to reconstruct outgoing long-wave radiation (OLR) under partial cloud conditions. OLR, measured in Top of the Atmosphere (TOA), was associated with integrating emissions from the ground, the lower atmosphere, and clouds [17]. Daily OLR data were used to study its variability in seismic activity zone [18,19]. Was suggested that the increase in radiation and the short-term change in OLR are associated with thermodynamic processes in the atmosphere over seismically active regions and are described as thermal radiation anomalies (TRA). The TRA characteristic was proposed in [20] as a maximum statistical change in OLR velocity for a particular spatial position and a predetermined time and was plotted following the anomalous thermal field [22,23]. Statistical relations between satellite thermal anomalies and major earthquakes have been studied extensively, and the connections have been established [24,25]. The first short-term warning tests based on Satellite Thermal anomalies was generated during in 2014 - 2015 over Japan [20]. Real-time forecast alerts (time, location, and magnitude) for M5.5+ were recorded in real-time over a period of 310 days, with 22 alarms being successfully issued with no false alarms.

## III. RESULTS

To illustrate the synergetic behavior of the SG process with the transient radiation anomalies in the atmosphere that occurred at a different stage of earthquake genesis, we explore two of the most significant earthquakes for the last 20 years: 2011 M9.0 Tohoku Earthquake, Japan, and M7.7/M7.3 2015 earthquakes in Nepal. We have chosen these earthquakes as they are the strongest and deadliest earthquakes for the last decade and also represent different geographic regions, different seismo-tectonic and meteorological conditions. The common between all the two cases is that none of those earthquakes have been officially forecasted, and the authors of the manuscript have reported warnings ahead of both events shared with the Disaster mitigation experts.

### 2011 Great Tohoku earthquake of March 11, Japan.

During the period of February 20 - 24, 2011, a powerful example of resonance vibration was reported in the Suez Canal. This is the same type of resonance that was reported in Russia (Volgograd and in Adygea) but on the surface of the water. It was assumed that a strong earthquake would occur on the opposite side of the Earth. This resonance was briefly reported on February 24, 2011, in a talk at the Science Museum in Haifa. On March 8, 2011, a major KaY wave anomaly was observed in Haifa (See Fig. 3A). The KaY wave was moving towards the East from the observations in Haifa (Israel), but with one sensor, it was challenging to provide an accurate forecast. Only a warning was issued about the observed anomaly and a trend about the direction of its movement towards Japan. On March 11, data appeared around a catastrophic earthquake in Japan. The initial satellite TRA post-event analysis has shown March 8-9 with anomalies pattern, over Tohoku earthquake [21]. The extended analysis based on comparison for the 11 years of analysis, including multiple National Oceanic and Atmospheric Administration (NOAA-15,18) satellites, revealed the most significant reading in the acceleration of OLR on the TOA much earlier on February 26 (Fig. 3B). The most significant TRA anomaly could indicate the geo-gas exhalation processes driven by the resonance vibration associated with the same seismo-gravity resonance. The additional acceleration in OLR near March 8-9 could be connected with degassing triggered by the KaY waves return.

### 2015 M.7.8 of Apr 24 and M7.3 of May 12, Nepal.

The authors did not forecast the primary M.7.8 earthquake of Apr 24, 2015, in Nepal due to insufficient ground stations (one only SGW station in operation) and lack of observation of operational satellite data. However, the two methods (SGW and TRA) were able to forecast the second large event of M7.2 in Nepal of May 12, 2015, days ahead, independently of each other. In order to help rescuers working in Nepal, a forecasting alert was sent to the first responders on the ground with the expected time of the earthquakes and approximate magnitude. SGW observation in Haifa detected an M7.2 KaY wave 48 hours in advance (Fig. 3 A/B). A forecast was sent to the head of the Russian rescue team (RRT) about a possible major earthquake on May 10-11. 2015. The approximate time is because only one sensor worked in Haifa during that period. The declared earthquake happened on May 12, 2015, and the head of the RRT sent a letter of thanks to A.Yagodin. We used OLR data from NOAA's Advanced Very High-Resolution Radiometer (AVHRR) for satellite thermal analysis over Nepal. The results of the 2015 Nepal earthquake showed that there was a rapid increase in transient infrared radiation in satellite data in mid-March 2015. An anomaly is observed near the epicenter; it peaked around April 4-7, about 2-3 weeks before M7.8. After the M.7.8 earthquake of Apr 24, 2015, further satellite analysis revealed another temporary OLR anomaly on May 3-4. An official forecast was released on May 3 ( 9 days in advance) that the second M7 + event would occur in Nepal. The second M7.3 event happened on May 12, 2015. [26]. Our results show that long-wave radiation signals associated with earthquake processes were observed near epicentral regions several days before the corresponding earthquakes. TRA hotspots appeared quickly, remained in the same regions for several hours, and then quickly dissipated

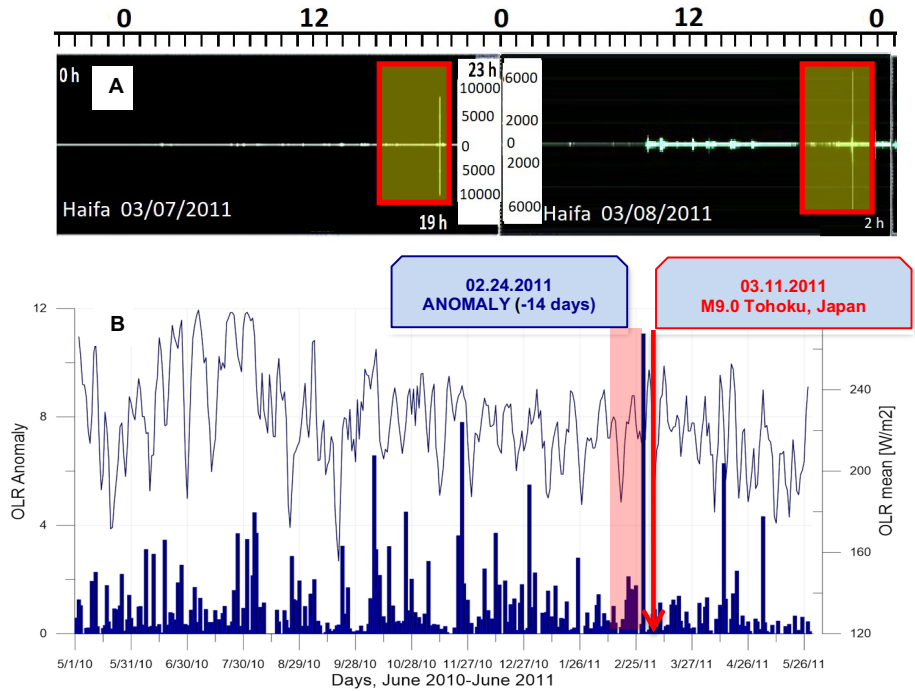

Fig.3 M9 2011 earthquake in Japan. A/ SG waveforms recorded in Haifa (Israel) on 02.24.2011 and 03.08.2011. KaY wave registered In Haifa, Israel with about 90 hours before the M9.0 in Japan of 03.11.2011. B/ TRA time series June 2010- june2011 over the Tohoku epicentral area. The largest TRA occurred on Feb 24, 2011, 15 day before the occurrence of M9.0 Tohoku earthquake.

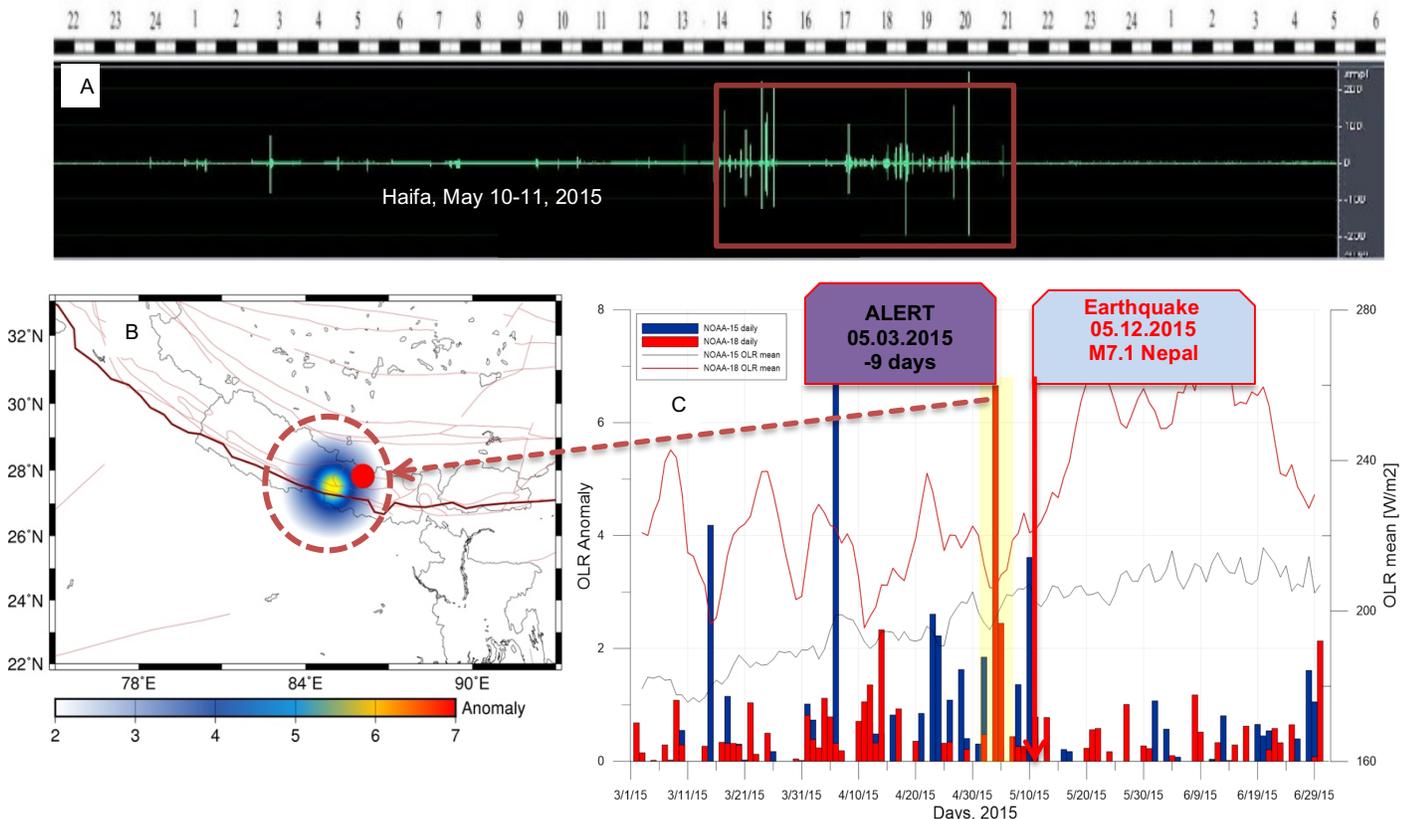

Fig.4 Nepal M earthquakes in 2015.A/ SG waveforms recorded in Haifa (Israel) 05.11.2015. KaY wave registered in Haifa, Israel with about 48 hours before the M7.3 in Nepal of 05.12.2015 B/ TRA time series for March 2015- June 2015 over the Nepal (Gorkha) epicentral area. The largest TRA's occurred on April 7, 17 days before the M7.8 of April 25, 2015, and on May 3, 2015, 9 days before the second Nepali earthquake of M7.3 of May 12, 2015. C/ TRA map of May 3, 2015. revealed by ongoing prospective analysis of satellite radiation revealed transient anomaly on May 3 (9 days in advance), associated with the M7.3 of May 12, 2015 earthquake.

5## IV. CONCLUSIONS

The largest earthquake in Japan for the last 70 years occurred near Sendai on March 11, 2011. SGW observations were able to indicate in advance for the 2011 Japanese EQ. On April 24 and May 12, 2015, M7.8 and M7.2 occurred in Nepal and brought major devastation to the region. Both SGW and Satellite observations, being independent of each other, we can alert in advance for the 2015 M7.3 Nepal earthquake of May 12. In addition, the two methods (SGW and TRA) were able to provide short-term forecasting during the test period, independently of each other, for the following notifiable earthquakes: M8.2 Okhotsk Sea 01.26.2013; M7.1 Mexico 09.19.2017; M7.6 Honduras of 01.18.2018; M6.8 Turkey of 01.24.2020 , M7.0 Dodecane Island of 10.30.2020, etc. This information was shared in advance with a very select audience and was not formally made available within the public domain. The joint analysis of the three major earthquakes M9, 2011 Japan; M7.8 and M7.2 in Nepal April 24 and May 12, 2015 demonstrates an evolution in the development of anomalous patterns that lead to forecasting solution. TRA anomalous data provide the temporal and spatial information of the initial seismo-gravity resonance process. SG waves in the form of KaY type of wave forecasted the time in hours and the direction to the future main earthquake event. A network of SGW sensors is necessary in order to provide the location on the future epicenter. One of the robust features of the SGW observations is that KaY-wave is repeatedly detected at the various located monitoring stations. If this waveform is visible all the way, then this gives an accuracy close to 100%. If the KaY type of waves is absent, then earthquakes are not happening for the entire observation period. Using Thermal radiation Anomalies methods, we generate up to 15 days of growing awareness of impending seismic activity based on satellite observation. KaY waves confirm the triggering of the physical earthquake rupture hours to days in advance, identifying the epicentral zone, short time window, and magnitude.

Acknowledgments

The authors thank NOAA's Climate Prediction Center (USA), for the OLR data. Special thanks go to the US Geological Survey and European-Mediterranean Seismological Centre for providing earthquake information services and data.Supplementary Material

The Supplementary Material for this article can be found online at: https://drive.google.com/drive/folders/1m5wcIshCrLSwQZi97dhAfcR9z25YKhI-

[1] Hayakawa M. and O. A. Molchanov (eds), (2002) Seismo Electromagnetics (Lithosphere - Atmosphere - Ionosphere Coupling), Terra Scientific Publishing, Tokyo, 420 p

[2] Hayakawa, M. (ed.), (2009) Electromagnetic Phenomena Associated with Earthquakes, Transworld Research Network, India, 279 pp.

[3] Hayakawa, M. (ed.) (2012) The Frontier of Earthquake Prediction Studies, Nihon- senmontosho-Shuppan, Tokyo, 794 pp.

[4] Pulinets S., K Boyarchuk (2004) Ionospheric precursors of earthquakes, Springer, 315pp.

[5] Pulinets, S. and Ouzounov, D. (2011), Lithosphere–atmosphere–ionosphere coupling (LAIC) model–a unified concept for earthquake precursors validation, J. Asian Earth Sci., 4, 371–382

[6] Ouzounov D., S. Pulinets, K.Hattori, P.Taylor (2018) (Ed's) "Pre-earthquake processes: A multi-disciplinary approach to earthquake prediction studies", 234, AGU/Wiley, 2018, 385p.

[7] Kozyrev N.A. (1972) On the Interaction between Tectonic Processes of the Earth and the Moon. In: Runcorn S.K., Urey H.C. (eds) The Moon. International Astronomical Union/Union Astronomique Internationale (Symposium No. 47 Held at the University of Newcastle-Upon-Tyne, England, 22–26 March, 1971), (1972) ,vol 47. Springer, Dordrecht. https://doi.org/10.1007/978-94-010-2861-5_21

[8] Yagodin A.(2017) Short-term forecasting of volcanic eruptions based on kay-wave monitoring, Austria-science,10, 14-16

[9] Yagodin A.(2018) Stages of the genesis of earthquakes, Austria-science, (2018b), 11,11-18 (in Russian)

[10] Yagodin A. , E.Yagodin (2018) Gravitational-seismic resonance - the basis of the genesis of earthquakes, Austria-science,12,11-19 (in Rus)

[11] Lutsenko E. A.Trunev N., N. Cherednychenko (2019) Resonant seismogenic and systemic-cognitive prediction of seismicity,256p, DOI 10,13140 / RG.2.2.18546.45760 (In Russian)

[12] Yagodin A. (2010) The exact parameters prediction of the future earthquake (a place, time, force) by measurement of the KaY-wave. Workshop on Earthquake Precursors, Tel Aviv University, Israel, 27 Jan2010,https://www.tau.ac.il/~colin/research/EarthQukes/Workshop/workshop3.html

[13] Yagodin A. (2018) Short-term earthquake prediction system, Scientific pages, 7, 3-10

[14] Pulinets S. and D. Ouzounov (2018) The Possibility of Earthquake Forecasting: Learning from nature, Institute of Physics Books, IOP Publishing, 168pp

[15] Gorny, V. I., Salman, A. G., Tronin, A. A., and Shilin, B. B. (1988), The Earth outgoing IR radiation as an indicator of seismic activity, Proc. Acad. Sci. USSR, 301, 67.

[16] Tronin, A. (2006), Remote sensing and earthquakes: a review, Phys. Chem. Earth, 31, 138-142.

[17] Ohring, G. and Gruber, A. (1982), Satellite radiation observations and climate theory, Adv. Geophys., 25, 237–304.

[18] Ouzounov D., D. Liu, C. Kang, G. Cervone, M. Kafatos, P. Taylor, (2007) Outgoing Long Wave Radiation Variability from IR Satellite Data Prior to Major Earthquakes , Tectonophysics, 431, 211- 220.

[19] Xiong P., X.Shen, Y. Bi, C.L. Kang, L.Z. Chen, F. Jing, Y. Chen (2010) Study of outgoing long wave radiation anomalies associated with Haiti earthquake, Nat. Hazards Earth Syst. Sci., 10, 2169-2178.

[20] Ouzounov D, S.Pulinets, Tiger Liu, K. Hattori, and P. Han (2018b) Multiparameter Assessment of Pre-Earthquake Atmospheric Signals; In the Book: Pre Earthquake Processes: A Multidisciplinary Approach to Earthquake Prediction Studies, Geoph. Monograph 234, John Wiley & Sons, 339-357

[21] Ouzounov D. S.Pulinets, K.Hattori, M, Kafatos, P.Taylor (2011) Atmospheric Signals Associated with Major Earthquakes. A MultiSensor Approach, in the book "Frontier of Earthquake short-term prediction study", M Hayakawa, (Ed), Japan, 510-531

[22] Tramutoli V., G. Di Bello, N. Pergola, S. Piscitelli (2001) Robust satellite techniques for remote sensing of seismic active areas, Ann. Geophys., 44, 295-312.

[23] Tramutoli V. (2015) From visual comparison to Robust Satellite Techniques: 30 years of thermal infrared satellite data analyzes for the study of earthquake preparation phases ,BGTA 56, 2, 167-202.

[24] Tramutoli, V., Filizzola, C., Genzano, N. and Lisi, M. (2018). Robust Satellite Techniques for Detecting Preseismic Thermal Anomalies. In the Book: Pre Earthquake Processes: A Multidisciplinary Approach to Earthquake Prediction Studies, Geoph. Monograph 234, John Wiley & Sons, https://doi.org/10.1002/9781119156949.ch14

[25] Genzano, N., Filizzola, C., Hattori, K., Pergola, N., & Tramutoli, V. (2021). Statistical correlation analysis between thermal infrared anomalies observed from MTSATs and large earthquakes occurred in Japan (2005–2015). Journal of Geophysical Research: Solid Earth, 126, e2020JB020108. https://doi.org/10.1029/2020JB020108

[26] Ouzounov, D., S. Pulinets, and D. Davidenko (2015) Revealing pre-earthquake signatures in atmosphere and ionosphere associated with 2015 M7.8 and M7.3 events in Nepal. Preliminary results, https://arxiv.org